\documentclass[sigconf,screen]{acmart}

\AtBeginDocument{%
  \providecommand\BibTeX{{%
    \normalfont B\kern-0.5em{\scshape i\kern-0.25em b}\kern-0.8em\TeX}}}

\setcopyright{acmcopyright}
\copyrightyear{2023}
\acmYear{2023}
\setcopyright{acmlicensed}\acmConference[ICTIR '23]{Proceedings of the 2023 ACM SIGIR International Conference on the Theory of Information Retrieval}{July 23, 2023}{Taipei, Taiwan}
\acmBooktitle{Proceedings of the 2023 ACM SIGIR International Conference on the Theory of Information Retrieval (ICTIR '23), July 23, 2023, Taipei, Taiwan}
\acmPrice{15.00}
\acmDOI{10.1145/3578337.3605132}
\acmISBN{979-8-4007-0073-6/23/07}

\usepackage{multirow}
\usepackage{subcaption}

\begin{document}

\title{Learn to be Fair without Labels: a Distribution-based Learning Framework for Fair Ranking}

\author{Fumian Chen}
\affiliation{%
  \institution{Institute for Financial Services Analytics\\
  University of Delaware}
  \city{Newark}
  \state{DE}
  \country{USA}
}
\email{fmchen@udel.edu}

\author{Hui Fang}
\affiliation{%
 \institution{Department of Electrical and Computer Engineering\\
 University of Delaware}
  \city{Newark}
  \state{DE}
  \country{USA}
}
\email{hfang@udel.edu}


\begin{abstract}
  Ranking algorithms as an essential component of retrieval systems have been constantly improved in previous studies, especially regarding relevance-based utilities. In recent years, more and more research attempts have been proposed regarding fairness in rankings due to increasing concerns about potential discrimination and the issue of echo chamber. These attempts include traditional score-based methods that allocate exposure resources to different groups using pre-defined scoring functions or selection strategies and learning-based methods that learn the scoring functions based on data samples. Learning-based models are more flexible and achieve better performance than traditional methods. However, most of the learning-based models were trained and tested on outdated datasets where fairness labels are barely available. State-of-art models utilize relevance-based utility scores as a substitute for the fairness labels to train their fairness-aware loss, where plugging in the substitution does not guarantee the minimum loss. This inconsistency challenges the model's accuracy and performance, especially when learning is achieved by gradient descent. Hence, we propose a distribution-based fair learning framework (DLF) that does not require labels by replacing the unavailable fairness labels with target fairness exposure distributions. Experimental studies on TREC fair ranking track dataset confirm that our proposed framework achieves better fairness performance while maintaining better control over the fairness-relevance trade-off than state-of-art fair ranking frameworks. 
\end{abstract}





\maketitle

\section{Introduction} \label{introduction}
Ranking algorithm in information retrieval (IR) systems is one of the most important determinants of how people consume information online. Over the past decades, researchers proposed countless ranking algorithms to maximize user utilities and improve search experiences. Without consideration of fairness, these utility-based algorithms might cultivate a vicious circle that makes search results dominated by majority groups and even polarizes the online community \cite{celis2019controlling}. 

Fair ranking in IR systems has been attracting more and more attention for long-term sustainability and to prevent potential discrimination. Related work regarding fair ranking methods can be divided into two groups, score-based or supervised learning-based  \cite{zehlike2021fairness}. Score-based methods achieve fair ranking by pre-defined scoring functions based on a desired score distribution, which are heavily content-dependent. The pre-defined scoring functions are less flexible and sometimes incapable of processing out-of-sample data points if statistically impossible \cite{le2022survey}. Therefore, many supervised learning-based fair-ranking algorithms have been proposed to learn the scoring functions for a more robust and data-driven solution. Although supervised-learning methods are proven to promote ranking fairness based on their evaluation metrics, they still face significant limitations. (1) Many algorithms were trained on outdated benchmark datasets, which are often unsuitable for ranking tasks. Besides, with numerical features only, contextual features based on text fields, which can potentially improve model performance and interpretability, are out of their scope \cite{patro2022fair}. (2) Lacking fairness gold labels, these supervised learning-based models use relevance or utility scores to substitute the fairness labels, which do not necessarily reflect fairness. Consequently, plugging the substitution as the gold label does not guarantee the minimum fairness-aware loss, and using gradient descent to train these fairness models is problematic.

In this study, we propose a distribution-based fair learning framework (DLF) to achieve fair ranking without labels and help address the limitations of existing learning-based fair-ranking algorithms. To train and test our model on state-of-art fair ranking datasets, we adopted the experimental dataset used by the TREC fair ranking track \footnote{\url{https://fair-trec.github.io/}} for a Wikipedia retrieval task. It is designed specifically for IR and ranking tasks comprising more than six million Wikipedia articles with full-text fields. This enables us to explore the value of underexploited contextual features using natural language processing (NLP) techniques. Given the difficulty of acquiring fairness annotations, we formulate a distribution-based and fairness-aware loss function that does not require fairness labels but can converge through gradient descent by replacing the fairness labels with a target fairness distribution, which reflects the fairness goal of a ranking. Last, detecting and managing fairness and relevance trade-offs has been a difficult task confirmed in previous studies \cite{gao2022fair, wang2021understanding, patro2022fair}. To ensure the final rankings are relevant and fair with good control over the trade-off, we separate fairness and relevance models first and then merge them with a weighted sum function. To the best of our knowledge, we proposed the first distribution-based fair learning framework without fairness labels, and we make the following contributions:
\begin{itemize}
\item We proposed a distribution-based fair learning (DLF) framework that (1) leverages contextual features to improve model performance, (2) does not require gold fairness labels, and (3) separates the fairness model from the relevance model and can ideally merge with any relevance model.
\item Our framework performs better than existing fair ranking frameworks, especially regarding fairness and managing the relevance-fairness trade-off.
\end{itemize}

\section{Related Work} \label{related_work}
Fair ranking algorithms can be divided into score-based and supervised learning-based methods \cite{zehlike2021fairness}. Both methods use scoring functions to rank, but score-based methods pre-define the functions, whereas supervised-learning methods train the functions based on training data.
Using score-based methods often requires interventions or enforces constraints on either score distributions or scoring functions to reduce unfairness \cite{yang2017measuring, yang2019balanced, celis2017ranking, stoyanovich2018online}. Recent score-based methods utilize statistical and selection models \cite{zehlike2017fa, zehlike2022fair} or greedy algorithms \cite{gao2022fair} to ensure items that maximize fairness utilities or fill the desired proportional representations are ranked at top positions. These score-based methods are often intuitive and interpretable but require task-specific inputs, such as desired percentages of protected groups or significance levels. Hence, they are heavily context-dependent and sometimes statistically infeasible, especially when multiple fairness groups are involved. 

Participants in the TREC fair ranking track also tested various techniques to produce fair rankings. Many of them utilize diversification-based methods, such as MMR \cite{cherumanal2021rmit}, PM-2 and rank fusion\cite{cherumanalrmit}, and heuristic approaches \cite{jaenich2021university}. Vardasbi et al. \cite{ali2021university} leveraged LamdaMART \cite{burges2010ranknet}, ListNET \cite{cao2007learning}, and Logistic regression to maximize evaluation metrics via swapping positions. There are also score-based methods that convert list-wise fairness measures to point-wise estimations and then build a model accordingly \cite{chenexploration, zhuoqitkb48}.

This study focuses on supervised-learning-based methods to develop a more flexible fair ranking algorithm. Previous supervised learning models encode fairness using pre-processing, in-processing, or post-processing strategies \cite{zehlike2021fairness}. Pre-processing \cite{lahoti2019ifair} and post-processing \cite{zehlike2020matching} methods modify either training or predicted data to ensure they are fairly represented. In-processing methods leverage fairness-aware losses or constraints and then solve optimization problems to achieve fair ranking. Zehlike et al. \cite{zehlike2020reducing} represented DELTR framework that incorporates a list-wise unfairness utility based on disparate exposure to their loss function:
\begin{equation} \label{deltr_loss}
    L_{DELTR}(y^{(q)}, \hat{y}^{(q)}) = \mathbb{E}(y^{(q)}, \hat{y}^{(q)}) + \gamma \mathbb{U}(\hat{y}^{(q)})
\end{equation}
where $y^{(q)}$ is the relevance gold label, $\mathbb{E}(y^{(q)}, \hat{y}^{(q)})$ is the cross-entropy loss, and $\mathbb{U}(\hat{y}^{(q)})$ is an unfairness measurement. Wang et al. \cite{wang2022meta} proposed a meta-learning-based fair ranking algorithm (MFR) to ensure a modified version of DELTR can be better trained on a biased dataset. These state-of-art learning models use disparate exposure, which measures the exposure difference between two groups, the protected and the unprotected groups, to construct their loss. Subsequently, they can only handle binary sub-groups. Even though many fairness evaluation frameworks have been proposed to handle multinary sub-groups regarding multiple fairness categories by measuring the distance between position-aware exposure distributions \cite{gao2022fair, sakai2022versatile}, we cannot simply apply their fairness measures to learning loss because these evaluation metrics are not differentiable and optimal rankings cannot be learned by gradient descent. On the other hand, existing learning-based models were often trained and tested on outdated benchmarks, such as the COMPAS dataset \cite{propublica} and the Engineering Students dataset \cite{milkalichtblau}, which are not designed for the retrieval and ranking tasks \cite{patro2022fair}. Being trained on outdated datasets where numerical features are the only available features and relevance is the only available label, learning-based methods like DELTR and MFR face severe limitations. Firstly, numerical features cannot well-represent documents, especially when full-text fields are available but underexploited, leading to poor out-of-domain performance and low interpretability. Moreover, since plugging in the relevance label does not guarantee the minimum loss, as shown in Eq. (\ref{deltr_loss}), using gradient descent methods to obtain the optimal weights is questionable. Another challenge of existing fair ranking algorithms is managing the fairness-relevance trade-off \cite{gao2022fair, wang2021understanding, gao2021addressing}, which has shown to be difficult to quantify and manage. Optimizing a loss containing both relevance and fairness components, existing fair ranking algorithms like DELTR have little control over the trade-off, and tuning their preference parameter over fairness is computationally costly. 

Therefore, we introduce the distribution-based fair learning framework without fairness labels to (1) fix the inconsistency between unavailable fairness labels and training loss, (2) improve model performance and interpretability using contextual features extracted from the full-text field, and (3) manage the fairness-relevance trade-off better by separating and merging fairness and relevance models.

\section{Fair Ranking Task Formulation} \label{task_form}

Let $Q$ be a set of queries, where each query $q_i$ consists of a set of documents $D =(d_1, d_2, ..., d_n)$ to be ranked. Each document $d_i$ is associated with a set of features $X_i=(x_1,x_2,...,x_j)$, and a relevance label $y_{i,j} \in \{0, 1\}$ indicating whether the document is relevant or not to each query $q_i$. Generally, we aim to find the best permutation (rank) $\pi_q$ consisting of the document set D for query $q$ that (1) meets the task's fairness definition  and (2) ranks relevant documents at higher positions.

Exposure distribution-based fairness definition is the most effective way and the consensus adopted by recent fairness evaluation frameworks \cite{gao2022fair, sakai2022versatile, raj2020comparing, sapiezynski2019quantifying}. An exposure distribution consists of the percentage of exposures each sub-group receives regarding a fairness category (e.g., gender). For example, if 50\% of exposure is given to the sub-group male, 40\% is given to the sub-group female, and the rest is given to the sub-group of non-binary, then the exposure distribution is: 
\begin{equation}
    \epsilon = (P_{\text{male}}=0.5, P_{\text{female}}=0.4, P_{\text{non-binary}}=0.1)
\end{equation}
The exposure distribution-based fairness definition calculates the distance between a system-produced exposure distribution $\epsilon(\pi)$ and the target exposure distribution $\epsilon^*$. Systems that produce a small distance between these two distributions would be considered fair.

More than one fairness category might be involved in fair ranking tasks in the real world. For example, an IR system might want to produce fair rankings regarding both gender and geographic locations at the same time. Let $G_{ij}$ be a set of fairness groups we are interested in to provide fair exposure, which consists of $i$ categories with $j$ sub-groups of each category. Each permutation (rank) of the document set $D$ forms exposure distribution $\epsilon$ based on the group membership of each document in the set. Now, we restate the goal of fair ranking: finding the best permutation (rank) $\pi_q$ consisting of the document set D for query $q$ that (1) provides an exposure distribution $\epsilon$ that is close to its target exposure distribution $\epsilon^*$ respecting one or more fairness categories, and (2) ranks relevant documents at higher positions.

Last, if the initial list of documents $D$ was given for each query, then we only need to re-rank the documents. Otherwise, a retrieval model is also needed to retrieve the initial ranking first.

\section{Distribution-based Fair Learning Framework without Labels} \label{iLTR}
As mentioned in Section \ref{introduction} and \ref{related_work}, the existing supervised learning-based fair ranking algorithms have two major limitations that weaken their reliability and might impair their model performance. In this section, we introduce our proposed framework and address these limitations.

\subsection{Distribution-based Fairness-aware Loss}
One limitation is that existing fair ranking frameworks use scores that are not necessarily related to fairness as their training ground truth label. To address this, we proposed the distribution-based fairness-aware loss, which does not require any fairness gold labels. Here, we demonstrate how the loss is constructed.

Given the fair ranking goal and fairness definition mentioned in Section \ref{task_form}, we use both target and system-produced exposure distribution to construct our fairness-aware loss. First, we discuss how to form the target exposure distribution. We adopted a well-accepted fairness definition that the exposure distribution across different groups should be equal or proportional to their utilities or impacts \cite{diaz2020evaluating}. That is, we estimate the target exposure distribution $\epsilon^*$ based on all relevant documents associated with a given query. For example, given a query where 50\% of all relevant documents belong to the "male" group, 40\% to the "female" group, and the remaining documents are marked as "non-binary," the target distribution is $  \epsilon^*_\text{gender} =(P_{\text{male}}=0.5, P_{\text{female}}=0.4, P_{\text{non-binary}}=0.1)$. Having the target distribution, we also need system-produced exposure distributions, which reflect how much exposure each sub-group received by the system. Evaluation frameworks utilize position-aware decay functions to construct system-produced exposure distribution. However, we cannot use the same method in our learning framework because loss with ranking and sorting are not differentiable and cannot be minimized through gradient descent.

The loss is, therefore, constructed inspired by the \textbf{top-one probability} \cite{cao2007learning}, which reflects the probability of a document $d_i$ with a score $s_i$ being ranked at the top position, such that:
\begin{equation}
    P(s_i) = \frac{\phi(s_i)}{\sum_{k=1}^n\phi(s_k)}
\end{equation}
where $\phi$ is a non-decreasing positive function, and $k$ is the length of the list to be rank. It is clear that given two objects $i$ and $j$ with $s_i > s_j$, then $P(s_i)>P(s_j)$, and given a list of $n$ documents, $\sum_1^n P(s_k) = 1$. Ranking based on this probability gives the best permutation.

Similarly, we assume there exists a gold score of fairness $s^*$ for each item in a ranking list reflecting the contribution of the item to ranking fairness. Then, we can construct the probability of a document $d_i$ with a score of fairness $s_i$ being ranked at the top position respecting its fairness contribution, and we call it \textbf{top one fair probability}:
\begin{equation}
    P_{\text{fair}}(s_i) = \frac{\phi(s_i)}{\sum_{k=1}^n\phi(s_k)}
\end{equation}
This probability shares the same property as the original top-one probability. Since given a list of $n$ items, $\sum_1^n P(s_k) = 1$, we assume the total exposure received by the list is one, and each item within the list received a portion exposure quantified by its top one fair probability. Then, we can construct the system-produced exposure distribution of one fairness category $i$ produced by any permutation $\pi$ consisting of $n$ documents as:
\begin{equation} \label{eps_est}
    \epsilon_i(\pi) = \sum_{k=1}^n P_{\text{fair}}(s_k)*GM_{ik}
\end{equation}
where $GM_{ik}$ is the group membership matrix w.r.t. fairness category $i$ of document $d_k$. For example, if a document belongs to the sub-group male, we have:
$$
GM_{ik} = \begin{bmatrix} x_{\text{male}}=1 \\ x_{\text{female}}=0 \\ x_{\text{non-binary}}=0 \end{bmatrix}
$$
where the values in the group membership matrix can also be percentages to indicate partial and soft membership of a document.
Combining with the target exposure distribution $\epsilon_i^*$ for the fairness category $i$, we construct fairness loss w.r.t fairness category $i$ given permutation $\pi$ as:
\begin{equation}
\textit{L}_i(\pi) = KL(\epsilon_i(\pi), \epsilon_i^*)
\end{equation}
where $KL$ is the Kullback–Leibler divergence \footnote{\url{https://en.wikipedia.org/wiki/Kullback-Leibler\_divergence}} which measures the distance between two distributions. \\
\textbf{Final Loss Function.} Given the need of providing fairness to multiple fairness categories (e.g., gender and geographic location) simultaneously. We incorporate $m$ fairness categories into the final loss function by computing the weighted sum loss of each category:
\begin{equation} \label{final_loss}
\begin{aligned}
\textit{FL}(\pi) &= \sum_{i=1}^m w_i * \textit{L}_i(\pi) \\
&= \sum_{i=1}^m w_i *  KL(\epsilon_i(\pi), \epsilon_i^*)
\end{aligned}
\end{equation}
Here, we set the same weights $w_i$ for different fairness categories for simplicity. In future studies, we can adjust these weights to examine whether one fairness category is more important. 

\subsection{Learning without Labels}
Let's denote our loss in supervised learning settings. Given the function to learn $f(X, \theta) \to s$, where $X$ is the training features, and $s$ is the predicted fairness score, Eq. (\ref{final_loss}) can be re-write as:
\begin{equation}
\begin{aligned}
\textit{FL}(\pi) &= \sum_{i=1}^m w_i * KL(\sum_{k=1}^n P_{\text{fair}}(s_k)*GM_{ik}, \sum_{k=1}^n P_{\text{fair}}(s^*_k)*GM_{ik}) \\
&= \sum_{i=1}^m w_i * KL(\sum_{k=1}^n P_{\text{fair}}(f(X_k, \theta))*GM_{ik}, \sum_{k=1}^n P_{\text{fair}}(s^*_k)*GM_{ik})
\end{aligned}
\end{equation}
where $s^*$ is the gold label of fairness associated with each item in the optimal permutation. Then, the learning goal is to find the optimal $\theta^*$ that minimizes the loss:
\begin{equation} \label{eq_arg}
\begin{aligned}
\theta^* &= \arg \min \textit{FL}(\pi) 
\end{aligned}
\end{equation}
Now, we explain how the loss can be trained without fairness labels $s^*_k$. Since every component of loss $FL(\pi)$ is differentiable, we solve Eq. (\ref{eq_arg}) through gradient descent. However, the gold label $s^*$ is not given in this case and is barely available in most fair-ranking datasets. According to Eq. (\ref{eps_est}), $\epsilon^*_i = \sum_{k=1}^n P_{\text{fair}}(s^*_k)*GM_{ik}$, even though the optimal fairness score $s^*_i$ was not given, we are able to train the function $f(X, \theta) \to s$ using the target distribution $\epsilon^*$:
\begin{equation} \label{eq_arg1}
\begin{aligned}
\theta^* &= \arg \min \textit{FL}(\pi) \\
& = \sum_{i=1}^m w_i * KL(\sum_{k=1}^n P_{\text{fair}}(f(X_k, \theta))*GM_{ik}, \epsilon^*_i) 
\end{aligned}
\end{equation}
As can be seen from Eq. (\ref{eq_arg1}), no gold label of fairness $s^*$ is needed, and we only need the target distribution. 

Compared with previous LTR-based fair ranking algorithms with fairness-aware loss functions, such as DELTR mentioned in Section \ref{related_work} and Eq.(\ref{deltr_loss}), where plugging in the gold label, such that $y = y^*$, does not guarantee the lowest loss value, ideally, if we can train a function such that $s=s^*$, we will have $FL(\pi^*) = 0$. This consistency between the loss function and the gold label makes our gradient descent-based training more effective and interpretable.  

\subsection{Combining Fairness and Relevance} \label{sec_finalrank}
So far, the learning process does not involve any relevance component, and the features used are tailored for fairness. We call the model trained at this point our fairness model $DLF = f(X, \theta^*)$, which generates a fairness score $F$ for each evaluation data point with its feature set $X$. Merging our fairness model with a relevance model is important to ensure the final ranking is fair and relevant. We start with the most popular relevance retrieval model, BM25 \cite{Robertson1993}, to retrieve the initial rankings and merge them with the fairness model. Ideally, our fairness model can merge with any retrieval model that produces a relevance score $R$. With the score of fairness $F$ and the score of relevance $R$ obtained from the fairness-only and relevance-only models, the final ranking is then constructed based on the final score: 
\begin{equation}
    \textit{final score} = (1 - \alpha) * R + \alpha * F
\end{equation}
where $\alpha$ is a fairness preference parameter. This framework enables us to visualize and manage the relevance-fairness trade-off better than the previous frameworks that optimize relevance and fairness simultaneously. Last, as mentioned in Section \ref{task_form}, if the initial set of documents $D$ for query $q$ was not given, we should use a relevance retrieval model to obtain the initial ranking first, and it can be the same as our relevance model.

\subsection{Contextual Features Extraction} \label{cont_fe}
The other limitation is that many existing fair ranking studies obtain the scoring function by optimizing fairness and relevance simultaneously with the same set of features, which are often numerical and leave the underexploited text field out-of-scope. Even though their frameworks can theoretically use different features for fairness and relevance, non-previous studies have been explored using tailored features for fairness and relevance separately. The training feature set $X$ plays a crucial role regarding model performance, model's predictive power, and model interpretability when using supervised learning algorithms to learn the scoring function $f(X, \theta)$ as shown in Eq. (\ref{eq_arg1}). Our framework leverages the text field to extract contextual features and separates the relevance and fairness models so that they can be trained respectively with their tailored features. Specifically, we extracted the contextual features out of the text field based on word embeddings to augment training features for the fairness model. These contextual features are extracted to utilize potential semantic relationships between query/document and fairness categories and help our algorithm better capture patterns from training data. For instance, given target exposure distributions vary by different queries, contextual features that are extracted to capture the relationship between query and fairness categories could be significantly valuable. Table \ref{tab:fes} summarizes the features extracted for model training. Many natural language process techniques could be used to extract contextual features from the text field. In this study, all embeddings are based on Sentence-BERT \cite{reimers2019sentence} \footnote{\url{https://www.sbert.net/}} with the pre-trained model 'all-mpnet-base-v2'. Since the embedding model can only take 512 tokens at most, we split the long text field and merge the results using Spacy, an industrial-strength
natural language
processing python package,\footnote{\url{https://spacy.io/}} with its pre-trained model \textit{en\_core\_web\_trf}. 

\begin{table*}[htbp]
\centering
\resizebox{0.95\textwidth}{!}{%
\begin{tabular}{c|c}
\hline
\textbf{Feature Name}        & \textbf{Description}                                                                 \\ \hline
\textit{bm-25}               & \multicolumn{1}{|l}{The BM25 \cite{Robertson1993} score of the query-document pair}                                            \\ \hline
\textit{q\_gender\_sim}      & \multicolumn{1}{|l}{The cosine similarity between query embeddings and gender embeddings}                 \\
\textit{d\_gender\_sim}      & \multicolumn{1}{|l}{The cosine similarity between document embeddings and gender embeddings}              \\
\textit{q\_geo\_loc\_sim}    & \multicolumn{1}{|l}{The cosine similarity between query embeddings and geographic location embeddings}    \\
\textit{d\_geo\_loc\_sim}    & \multicolumn{1}{|l}{The cosine similarity between document embeddings and geographic location embeddings} \\ \hline
\textit{q\_gender\_sub\_sim} & \multicolumn{1}{|l}{The cosine similarity between query embeddings and gender sub-groups embeddings}      \\
\textit{d\_gender\_sub\_sim} & \multicolumn{1}{|l}{The cosine similarity between document embeddings and gender sub-groups embeddings}   \\
\textit{q\_geo\_sub\_sim} & \multicolumn{1}{|l}{The cosine similarity between query embeddings and geographic location sub-groups embeddings}    \\
\textit{d\_geo\_sub\_sim} & \multicolumn{1}{|l}{The cosine similarity between document embeddings and geographic location sub-groups embeddings} \\ \hline
\end{tabular}%
}
\caption{Summary of Training Features X for Fairness Model. Notice that the last four rows listed in the table are four groups of features: the cosine similarity between query/document embeddings and every sub-group (e.g., male, female) embedding, respectively. For example, \textit{q\_gender\_sub\_sim} is actually four features: \textit{q\_male\_sim}, \textit{q\_female\_sim}, \textit{q\_non-binary\_sim}, and \textit{q\_unknown\_sim}, given \textit{gender} has four sub-groups.}
\label{tab:fes}
\end{table*}

\section{Experiments}

\subsection{Experiment Settings}
\subsubsection{Experimental Dataset} \label{exp_datasets} We train and evaluate our proposed framework as well as other baselines on the Wikipedia dataset used by TREC fair ranking tracks \cite{ekstrand2023overview} and designed for fair ranking tasks. The corpus contains over six million English Wikipedia articles with a full-text field associated with 50 training queries from various domains and 50 evaluation queries. TREC provides binary relevance labels and the fairness metadata of each article. The fairness metadata contains fairness annotations regarding nine different fairness categories (i.e., geographic location, gender, age of the topic, popularity, etc.), enabling us to examine fairness w.r.t. multiple fairness categories. For simplicity, we start with the two most common categories, geographic location, and gender. This also aligns with our contextual features extraction, as mentioned in Section \ref{cont_fe}, and will be used for fairness evaluation.

\subsubsection{Evaluation Metrics} We adopt the same evaluation framework used by TREC. We use the attention-weighted ranking fairness (AWRF) \cite{sapiezynski2019quantifying, raj2020comparing} to measure fairness in rankings, which captures the statistical parity between cumulative exposure and a population estimator, such that:
\begin{equation}
    \text{AWRF}(\pi) = \Delta(\epsilon(\pi), \hat{\epsilon})
\end{equation}
where $\hat{\epsilon}$ is the population estimator or the target exposure and $\epsilon(\pi)$ is the cumulative and attention-weighted exposure generated by permutation $\pi$. A log-decay function is employed to show that exposure is position-aware. AWRF allows soft group membership with multiple fairness categories by employing the KL divergence to calculate distances \cite{raj2020comparing}. As mentioned in Section \ref{exp_datasets}, we start with two fairness categories to examine fairness w.r.t multiple fairness categories: \textit{gender} and \textit{geographic location}, where \textit{gender} has four sub-groups (e.g., male, female) and \textit{geographic location} has 21 sub-groups (e.g., North America, Asia). Relevance is evaluated by nDCG \footnote{\url{https://en.wikipedia.org/wiki/Discounted\_cumulative\_gain}}, a popular and effective relevance evaluation metric.

\subsubsection{Baseline Models} Our baseline models include (1) a relevance-only retrieval model BM25 \cite{Robertson1993}, (2) a random re-ranker that randomly ranks each item retrieved, (3) a diversification-based model, MMR (Maximal Marginal Relevance) \cite{carbonell1998use}, that diversifies the retrieved item list, (4) a score-based statistical model, the FA*IR \cite{zehlike2017fa} \footnote{\url{https://github.com/fair-search/fairsearch-fair-python}} model that post-processes the retrieved list so that items from protected groups meet a minimum percentage at top-ranking positions. (5) a state-of-art gradient-boosted tree (GBDT) model, LambdaMART \cite{burges2010ranknet},  and last, (6) the DELTR \cite{zehlike2020reducing} \footnote{\url{https://github.com/fair-search/fairsearch-deltr-python}} model with two different $\gamma$. 

As all models are built on retrieval and re-rank, we acquire the initial ranking with top 500 documents per query utilizing the BM25 \cite{Robertson1993} model implemented by Pyserini \footnote{\url{https://github.com/castorini/pyserini}}, a Python toolkit for reproducible information retrieval research \cite{Lin_etal_SIGIR2021_Pyserini}. For most of the baseline models, we adopted their default settings. We set $\lambda=0.5$ for the MMR model and use \textit{TfidfVectorier} \footnote{\url{https://scikit-learn.org/stable/modules/generated/sklearn.feature\_extraction.text.TfidfVectorizer.html}} to vectorize the text field and obtain the similarity matrix. Our LambdaMART model was implemented by LightGBM \footnote{\url{https://lightgbm.readthedocs.io/en/v3.3.2/}} with all default parameters for \textit{lightgbm.LGBMRanker}. For the DELTR model, we run two experiments with $\gamma_{small}=1$ and $\gamma_{large}=500$ for 500 iterations. For the FA*IR model, we set $k=400$, $p=0.3$, and $\alpha=0.15$. For a fair comparison, every LTR-based model we tested takes the same input feature set X, as shown in Table \ref{tab:fes}, and a binary relevance label y.

\subsubsection{DLF Settings} As discussed in Section \ref{iLTR}, an estimation of the target exposure distribution is needed for our DLF model. We estimate the target exposure distribution based on all relevant documents associated with a given query. Therefore, we combine all the relevant documents in the training set annotated by TREC for each query to construct the target exposure distributions concerning the two evaluation fairness categories: gender and geographic location. Finally, we leverage a multi-layer perceptron (MLP) model using PyTorch \footnote{\url{https://pytorch.org/}} to optimize our distribution-based fairness-aware loss and train the DLF model. We adjust the preference parameter $\alpha$ for the best combination of fairness and relevance scores. 

\subsection{Fairness Performance Analysis} \label{fair_res}

In Table \ref{tab:res_fair_frame}, we report DLF's fairness performance compared with existing fair ranking frameworks. Compared with the state-of-art fair ranking framework, DELTR, our DLF model achieves a higher AWRF score, regardless of small or large DELTR's fairness preference parameter $\gamma$. And the performance differences are statistically significant in most cases using the paired t-test based on 50 evaluation queries. Since DELTR utilizes relevance utility as their training label, plugging the label into their fairness-aware loss does not produce the minimum loss. Optimizing the loss by gradient descent is problematic and impairs the model performance. For evaluation queries (2022), trained solely on the relevance label, the DELTR model even failed to outperform the initial ranking. The direct comparison between our model and the DELTR model confirms the advantages of our proposed distribution-based learning framework. Our framework learned the scoring function more effectively by replacing the unavailable point-wise fairness labels with target-wise distributions to optimize the loss through gradient descent, as shown in Section \ref{iLTR}. DLF also outperformed another fair learning framework, the FA*IR. This is unsurprising because this score-based method cannot handle different queries as their scoring functions are pre-defined and fixed for all queries. Therefore, even though score-based methods do not require gold labels as well, they perform worse than LTR-based methods, especially when out-of-sample queries are from various domains and cannot be fitted into a fixed proportion of protected candidates in top positions.

\begin{table}[htbp]
\centering
\resizebox{\linewidth}{!}{%
\begin{tabular}{c|c|c}
\hline
 &
  \textbf{\begin{tabular}[c]{@{}c@{}}Evaluation Query (2021)\\ Fairness (AWRF@20)\end{tabular}} &
  \textbf{\begin{tabular}[c]{@{}c@{}}Evaluation Query (2022)\\ Fairness (AWRF@20)\end{tabular}} \\ \hline
\textit{Initial Ranking} & 0.6492          & 0.7217          \\ \hline
\textit{FA*IR \cite{zehlike2017fa}} & 0.6248          & 0.7237          \\
\textit{$\text{DELTR}_\text{small}$ \cite{zehlike2020reducing}} & 0.6530          & 0.6998          \\
\textit{$\text{DELTR}_\text{large}$ \cite{zehlike2020reducing}} & 0.6825          & 0.7202          \\ \hline
\textit{\textbf{DLF}}   & $\textbf{0.7501}^{\P \S \dag \ddag}$ & $\textbf{0.7402}^{\P \dag \ddag}$ \\ \hline
\end{tabular}%
}
\caption{Fairness performance of DLF compared with existing fair ranking frameworks. We report the AWRF@20 for evaluation queries from both years regarding two fairness categories, \textit{gender} and \textit{geographic location}. $\P$,$\S$,$\dag$, and $\ddag$ indicate DLF's statistically significant better performance (paired t-test based on 50 evaluation queries with p-value<0.05) over the initial ranking, FA*IR, $\text{DELTR}_\text{small}$, and $\text{DELTR}_\text{large}$ respectively.} 
\label{tab:res_fair_frame}
\end{table}


\subsection{Fairness and Relevance Performance Analysis}

\begin{table*}[htbp]
\centering
\resizebox{\textwidth}{!}{%
\begin{tabular}{c|ccc|ccc}
\hline
\multirow{2}{*}{} &
  \multicolumn{3}{c|}{\textbf{Evaluation Queries (2021)}} &
  \multicolumn{3}{c}{\textbf{Evaluation Queries (2022)}} \\ \cline{2-7} 
 &
  \multicolumn{1}{c|}{\textbf{Fairness (AWRF@20)}} &
  \multicolumn{1}{c|}{\textbf{Relevance (nDCG@20)}} &
  \textbf{Score (AWRF*nDCG)} &
  \multicolumn{1}{c|}{\textbf{Fairness (AWRF@20)}} &
  \multicolumn{1}{c|}{\textbf{Relevance (nDCG@20)}} &
  \textbf{Score (AWRF*nDCG)} \\ \hline
\textit{BM25 \cite{Robertson1993} Retrieval} &
  \multicolumn{1}{c|}{0.6492} &
  \multicolumn{1}{c|}{\textbf{0.2016}} &
  0.1308 &
  \multicolumn{1}{c|}{0.7217} &
  \multicolumn{1}{c|}{0.2502} &
  0.1805 \\
\textit{Random Re-ranker} &
  \multicolumn{1}{c|}{0.6754} &
  \multicolumn{1}{c|}{0.1155} &
  0.0780 &
  \multicolumn{1}{c|}{0.7280} &
  \multicolumn{1}{c|}{0.2249} &
  0.1637 \\
\textit{MMR \cite{carbonell1998use}} &
  \multicolumn{1}{c|}{0.6781} &
  \multicolumn{1}{c|}{0.1623} &
  0.1101 &
  \multicolumn{1}{c|}{0.7203} &
  \multicolumn{1}{c|}{0.2504} &
  0.1803 \\
\textit{LambdaMART \cite{burges2010ranknet}} &
  \multicolumn{1}{c|}{0.6556} &
  \multicolumn{1}{c|}{0.1231} &
  0.0807 &
  \multicolumn{1}{c|}{0.7105} &
  \multicolumn{1}{c|}{0.2141} &
  0.1521 \\ \hline
\textit{$\text{DELTR}_\text{large}$ \cite{zehlike2020reducing}} &
  \multicolumn{1}{c|}{0.6825} &
  \multicolumn{1}{c|}{0.1686} &
  0.1151 &
  \multicolumn{1}{c|}{0.7202} &
  \multicolumn{1}{c|}{0.2388} &
  0.1720 \\
\textit{$\text{DELTR}_\text{small}$ \cite{zehlike2020reducing}} &
  \multicolumn{1}{c|}{0.6530} &
  \multicolumn{1}{c|}{0.1376} &
  0.0899 &
  \multicolumn{1}{c|}{0.6998} &
  \multicolumn{1}{c|}{0.2340} &
  0.1637 \\
\textit{FA*IR \cite{zehlike2017fa}} &
  \multicolumn{1}{c|}{0.6248} &
  \multicolumn{1}{c|}{0.1834} &
  0.1146 &
  \multicolumn{1}{c|}{0.7237} &
  \multicolumn{1}{c|}{\textbf{0.2511}} &
  0.1817 \\ \hline
\textit{\textbf{DLF+BM25}} &
  \multicolumn{1}{c|}{$\textbf{0.7045}^{\P \S \dag \ddag}$} &
  \multicolumn{1}{c|}{$0.1923^{\dag \ddag}$} &
  $\textbf{0.1355}^{\S \dag \ddag}$ &
  \multicolumn{1}{c|}{$\textbf{0.7313}^{\P \dag \ddag}$} &
  \multicolumn{1}{c|}{$0.2507^{\ddag}$} &
  $\textbf{0.1834}^{\dag \ddag}$ \\\hline
\end{tabular}%
}
\caption{Fairness and relevance combined results. Since top positions receive the most attention in rankings, we report AWRF@20, nDCG@20, and a combined score (AWRF@20*nDCG@20) for both the evaluation queries (2021) and the evaluation queries (2022). The official evaluation metric for 2022 is @500, but we report @20 to make the scores for both years on similar scales. For the final ranking using our DLF model with BM25, we select the value of $\alpha=0.2$. Bold text indicates the best performance score. $\P$,$\S$,$\dag$, and $\ddag$ indicate DLF's statistically significant better performance (paired t-test based on 50 evaluation queries with p-value<0.05) over the initial BM25 Retrieval, FA*IR, $\text{DELTR}_\text{small}$, and $\text{DELTR}_\text{large}$ respectively.}
\label{tab:res-table}
\end{table*}

This section tested DLF's performance when merging with a relevance model compared with our baseline fair ranking algorithms. In Table \ref{tab:res-table}, we report model performance regarding the fairness metric, AWRF@20, relevance metric, nDCG@20, and a combined score, AWRF@20*nDCG@20. We divided the baseline models into two groups, relevance models and fair ranking frameworks. Our model (with $\alpha=0.2$) achieves the best performance regarding fairness and the combined score, as shown in the table. A paired t-test based on 50 evaluation queries proves the performance gaps are statistically significant.

The table shows that the DLF+BM25 model can improve the initial ranking's fairness and maintain a reasonable relevance score by comparing the first and last rows. For the evaluation query (2021), the existing fair ranking framework DELTR can also improve the fairness of initial ranking but fail to improve the evaluation query (2022). Our proposed model achieved a better fairness and relevance score in both sets of evaluation queries than the state-of-art fair ranking frameworks. DLF merging with BM25 \cite{Robertson1993} also outperformed the state-of-art GBDT-based model, LambdaMART. Since no fairness component is included, it is unsurprising that the model does not improve fairness, but at the same time, no improvements in relevance are observed either. This observation supports that training features are important, as mentioned in Section \ref{cont_fe}, and we should avoid using the same set of features for both relevance and fairness models. Features extracted specifically for fairness purposes are less likely to help relevance and vice versa. The proposed model also outperformed the random re-ranker and the diversification-based MMR model used by other groups participating in TREC. Regarding diversification-based methods, the results show that text diversification does not guarantee fair exposure across groups, and a more detailed examination is needed for future work.

\subsection{Additional Analysis}
This section discusses some additional analyses we did, including the value of contextual features analysis, convergence analysis, parameter sensitivity analysis, and the impact of initial ranking analysis.
\subsubsection{Value of Contextual Features and Convergence Analysis}

 \begin{figure}[htbp]
  \centering
  \begin{subfigure}[b]{0.45\textwidth}
    \includegraphics[width=\textwidth]{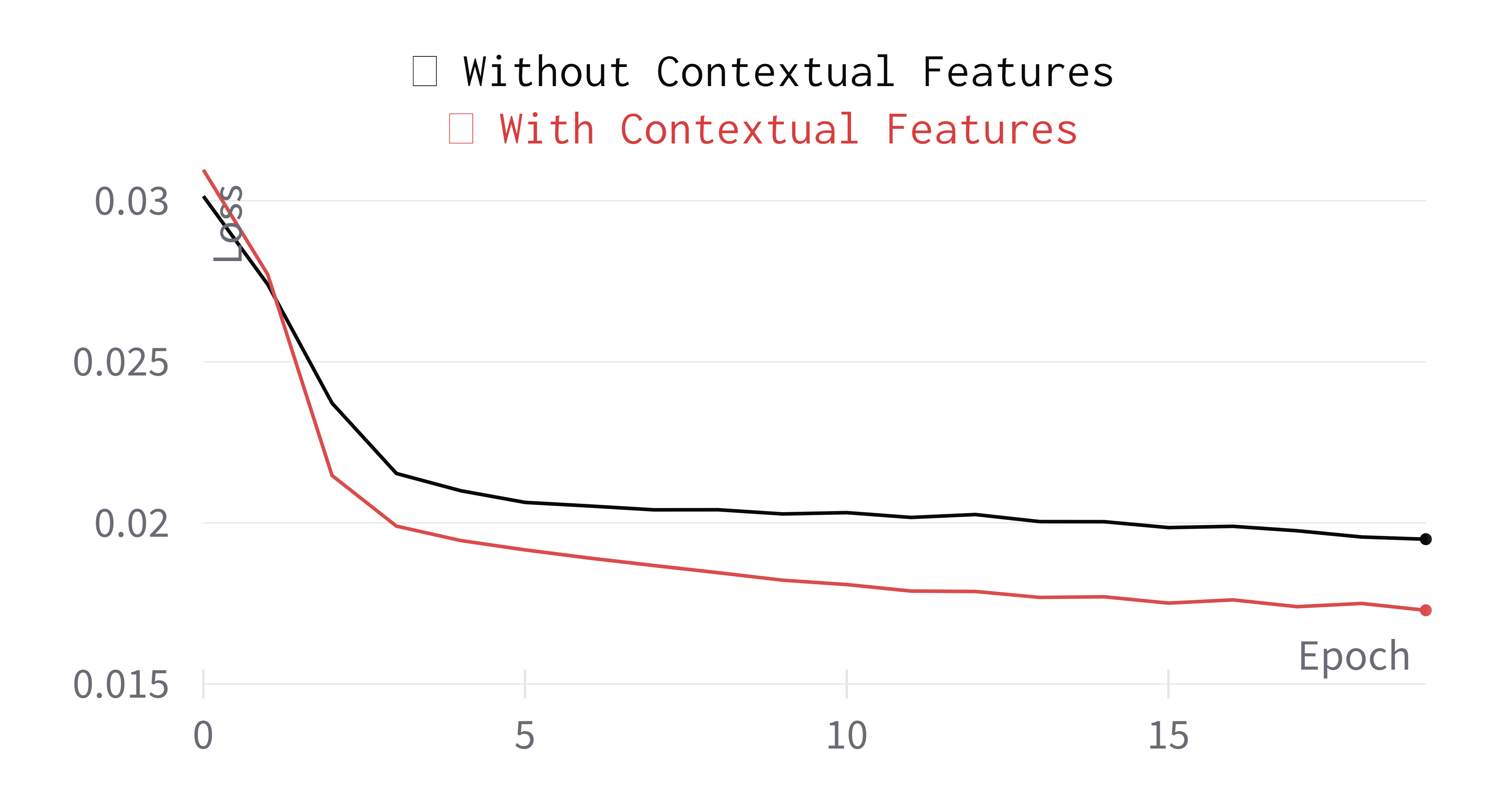}
    \caption{In sample training}
    \label{fig:insloss}
  \end{subfigure}
  \hfill
  \begin{subfigure}[b]{0.45\textwidth}
    \includegraphics[width=\textwidth]{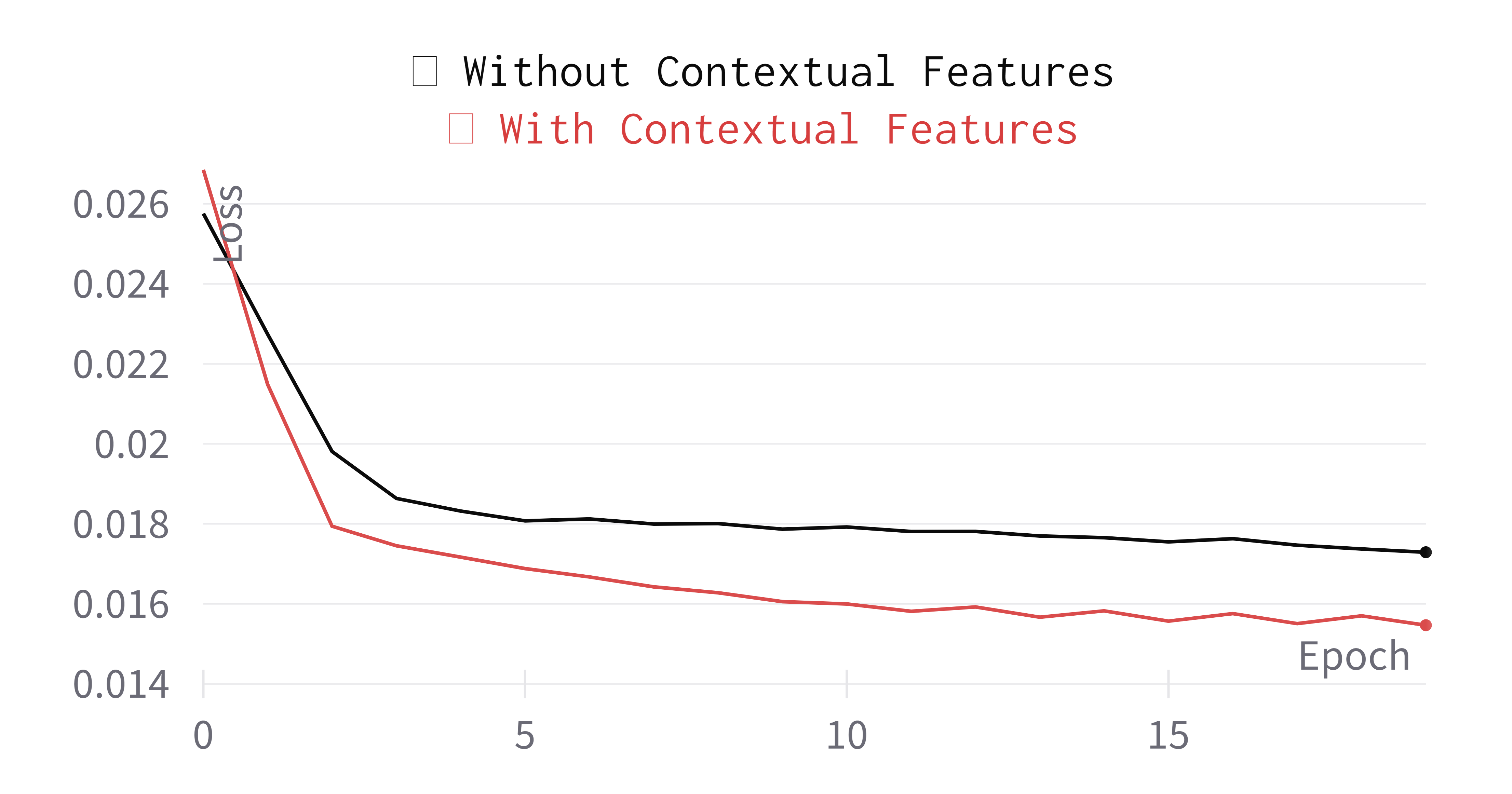}
    \caption{Out of sample validating}
    \label{fig:outsloss}
  \end{subfigure}
  \caption{Value of contextual features and convergence analysis plot. We plot the proposed distribution-based fairness-aware loss during training using gradient descent for 20 epochs with a learning rate = $10e-3$. Two sets of features were used to train the proposed model separately, one with contextual features in red and the other without contextual features in black.}
  \label{fig:loss_congverge}
\end{figure}

We first investigate the value of contextual features and the convergence nature of our proposed loss. According to Figure \ref{fig:loss_congverge}, we can first confirm that our proposed loss converges through gradient descent, and optimal weights are obtained. Second, by training our model based on two sets of features, one with contextual features and the other without, we can see that by using contextual features, the model achieves a lower loss and converge better than using the feature set without contextual features. This observation aligns with our motivation to better capture patterns in the experimental dataset using the underexploited text field to extract contextual features. Unlike previous fair ranking frameworks, which were trained on numerical features only, we tested a new direction to extract features for learning-based ranking models. This framework is particularly useful when constructing tailored features for different training purposes and when constructed numerical features are unavailable. Our DLF model, for example, is proposed for fair ranking. We extracted contextual features reflecting the relationship between document and fairness annotations to better capture patterns from the experimental dataset and to improve the model's performance and interpretability.

\subsubsection{Parameter $\alpha$ Sensitivity Analysis}

\begin{figure*}[!htbp]
  \centering
  \begin{subfigure}[b]{0.47\textwidth}
    \includegraphics[width=\textwidth]{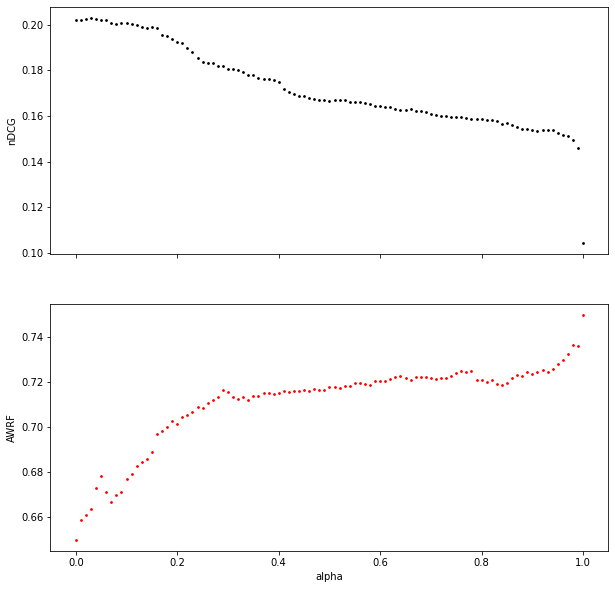}
    \caption{Evaluation Queries (2021)}
    \label{fig:frtrade1}
  \end{subfigure}
  \hfill
  \begin{subfigure}[b]{0.47\textwidth}
    \includegraphics[width=\textwidth]{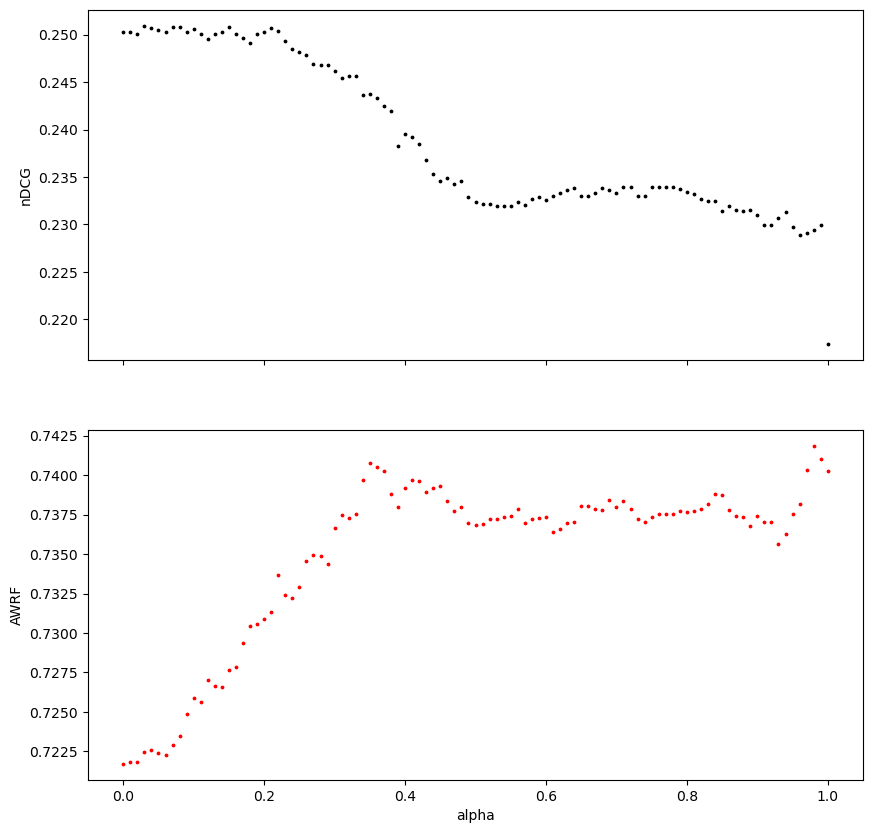}
    \caption{Evaluation Queries (2022)}
    \label{fig:frtrade2}
  \end{subfigure}
  \caption{DLF+BM25: Fairness and relevance plot by $\alpha \in [0,1]$ with an interval of 0.01. Generally, increasing $\alpha$ results in higher fairness (AWRF) and lower relevance (nDCG), but the increasing/decreasing is not linear and varies by different queries.}
  \label{fig:frtrade}
\end{figure*}

\begin{figure}[!htbp]
  \centering
  \includegraphics[width=\linewidth]{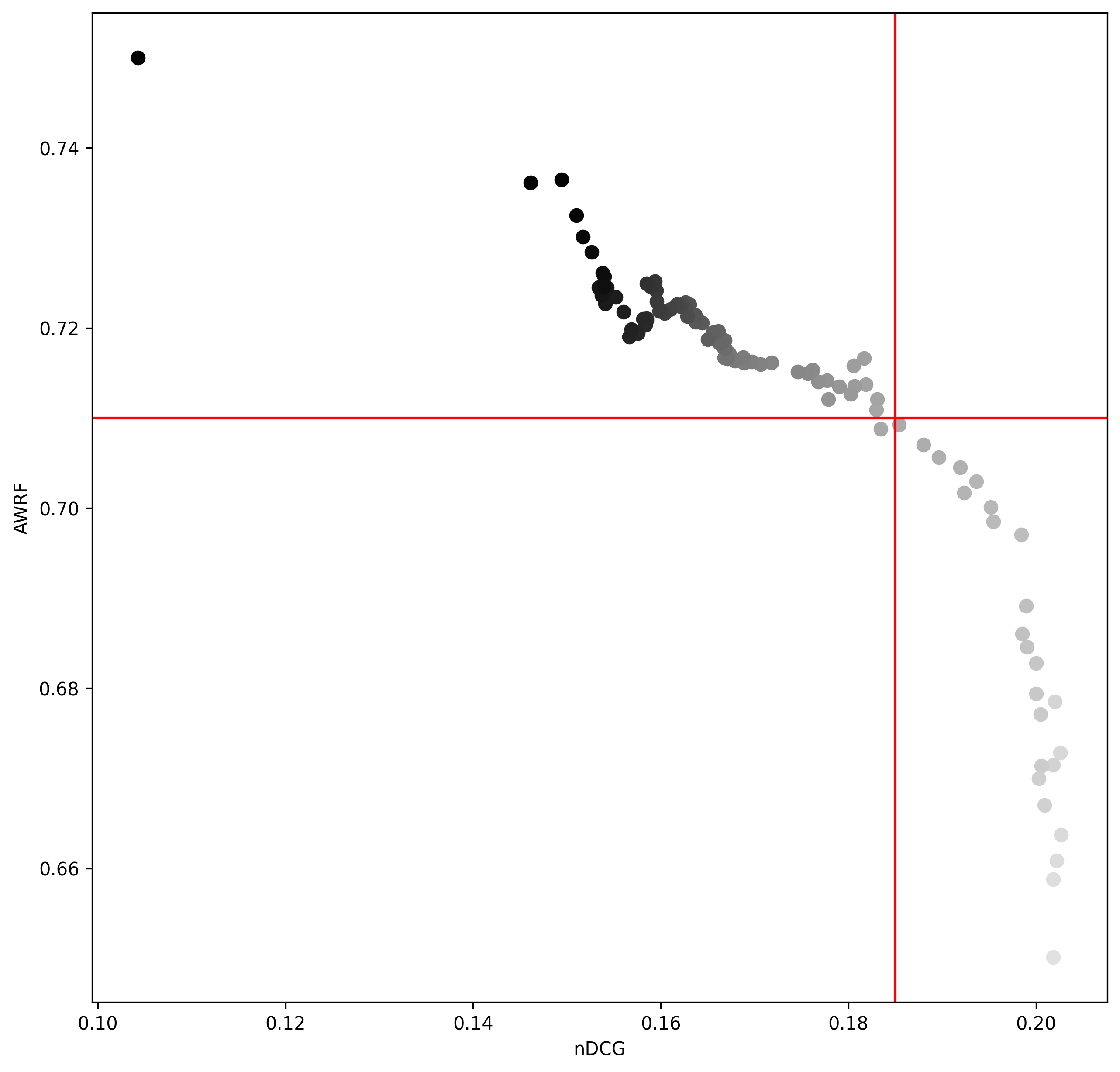}
  \caption{DLF+BM25: Fairness-Relevance trade-off plot (Evaluation Queries 2021) by the preference parameter $\alpha$. Darker dots indicate larger $\alpha$. Values of $\alpha$ are from 0 to 1 with an interval of 0.01. The large gap on the upper left implies the existence of documents that contribute to fairness but dramatically harm relevance. It also shows the difficulty of managing the trade-off.}
  \label{final-rank}
\end{figure}

We plot fairness and relevance separately by $\alpha \in [0,1]$ with an interval of 0.01 shown in Figure \ref{fig:frtrade} to examine how the selection of $\alpha$ impacts the final ranking. According to the plots, increasing the value of $\alpha$ generally increases the fairness score (AWRF) but decreases the relevance score (nDCG). However, the relationship between this increase/decrease and $\alpha$ is not linear, which encourages fair ranking algorithms using weighted sum functions for relevance and fairness to choose the value of $\alpha$ carefully. By adjusting $\alpha$ and visualizing the results, as shown in Figure \ref{final-rank}, our proposed model manages the relevance-fairness trade-off better than DELTR, where fairness and relevance are optimized together. When fairness and relevance are trained simultaneously, tuning models like DELTR for the best relevance-fairness combination is computationally costly. In our case, the best way to select parameter $\alpha$ is through visualization because the final ranking might be very sensitive regarding a small change of $\alpha$ and vice versa. That is, fairness can be improved with a minimum sacrifice of relevance, but a little improvement over fairness can also significantly damage relevance. Another advantage of employing our proposed model is that by training fairness and relevance separately, we can choose the optimal $\alpha$ based on the task context and how much we want to sacrifice relevance. For example, when users are searching for movies, retrieval systems might want to output more fair results to cover diverse movies from various sub-groups, even though relevance is compromised. In other general IR systems where users value relevance the most, fair results with a minimal loss of relevance are preferred.

\subsubsection{The Impact of Initial Ranking List}

\begin{figure}[htbp]
  \centering
  \includegraphics[width=\linewidth]{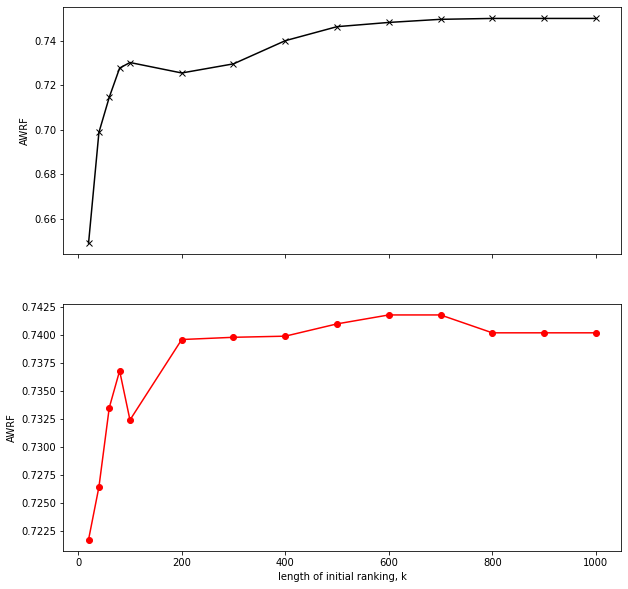}
  \caption{Fairness performance of DLF by re-ranking different lengths of initial rankings based on the top $k \in [20,1000]$ positions retrieved by BM25 \cite{Robertson1993}. AWRF@20 is used to plot the fairness boundaries by initial rankings with different lengths. The top line in black is for the evaluation query (2021), and the bottom line in red is for the evaluation query (2022).}
  \label{initial_rank_dlf}
\end{figure}

This subsection explored the impact of initial rankings for the proposed fairness-aware model before merging with a relevance model. Particularly, we tested two factors that determine the quality of initial rankings, the length of initial rankings and the retrieval models used to generate the initial rankings.

Regarding the length of initial rankings, we selected fourteen different lengths of initial rankings based on the top $k \in [20,1000]$ position at the initial ranking retrieved by BM25 \cite{Robertson1993}. Then, we plot the fairness boundaries (reported as AWRF@20) by initial rankings with different lengths in Figure \ref{initial_rank_dlf}. As can be seen, including more candidates in initial rankings helps our model achieve better fairness scores. This is intuitive because more candidates bring more possibilities of permutations to produce more fair rankings. However, as long as the initial ranking contains enough candidates, including more candidates does not always help. From Figure \ref{initial_rank_dlf}, we observed that the line converges to a straight line or even decreases when the length of initial rankings reaches 900. This implies that the length of initial ranking matters for re-ranking-based fair ranking algorithms. And the length of the initial ranking also impacts the effectiveness of our DLF model. 

\begin{figure}[htbp]
  \centering
  \includegraphics[width=\linewidth]{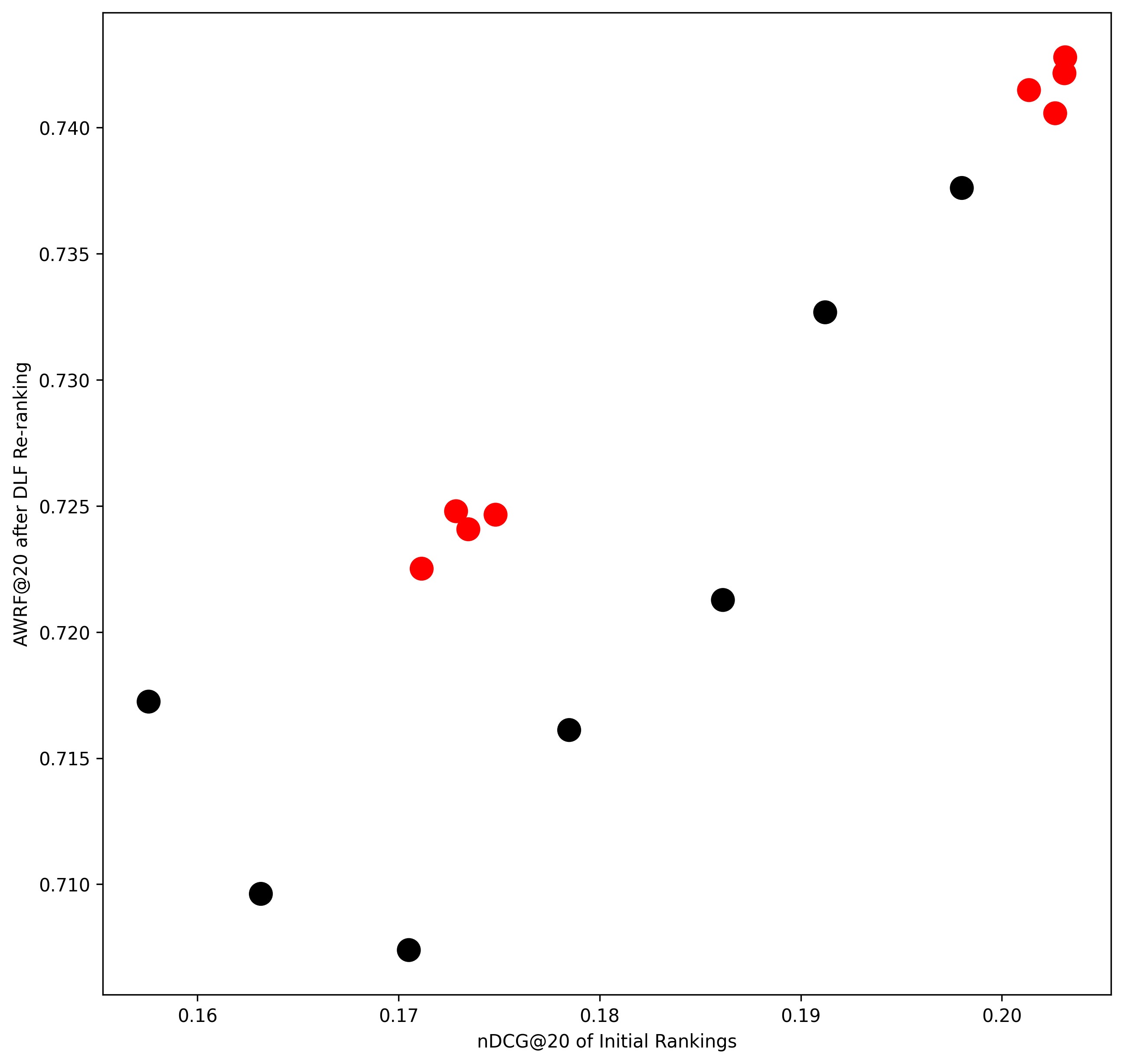}
  \caption{The relationship between relevance of initial rankings and fairness performance of DLF. We selected two different models, BM25 \cite{Robertson1993} (in Red) and RM3 \cite{lv2009comparative} (in Black), with different retrieval parameters to obtain the samples of initial rankings.}
  \label{initial_rank_retrieval}
\end{figure}

To examine the impact of using different retrieval models to obtain the initial rankings, we select two different retrieval models, BM25 \cite{Robertson1993} and RM3 \cite{lv2009comparative}, with different retrieval parameters, to construct different initial rankings having different nDCG@20 scores. Then, we re-rank these initial rankings with our DLF model and compute the AWRF@20 scores. We plot the results in Figure \ref{initial_rank_retrieval}. Initial rankings retrieved from BM25 \cite{Robertson1993} are marked in red, and initial rankings retrieved from RM3 \cite{lv2009comparative} are marked in Black. According to the plot, DLF's fairness performance highly correlates to the initial ranking's relevance. Good quality of initial rankings with higher relevance scores helps DLF to achieve better fairness performance. It is worth exploring more complex retrieval models that bring different re-ranking candidates. However, given the GPU memory limitation, we could not replicate the retrieval model, ColBERT-PRF \cite{wang2023colbert}, used by one of the TREC participants \cite{jaenich2021university} that produces the best TREC evaluation scores. We leave this part as one direction of our feature work. But based on the positive correlation observed in Figure 5, we expect the DLF model can bring even better fairness performance for a better retriever such as ColBERT-PRF.

\section{Conlusion and Future Work}

In this study, we proposed a distribution-based fair learning framework, which does not require fairness gold labels by replacing point-wise fairness labels with target-wise fairness exposure distributions. This opens a new problem-solving direction for future learning-based fair ranking studies when point-wise fairness ground truth is unavailable. Experiments conducted on the Wikipedia dataset used by TREC fair ranking track prove that our framework outperforms existing fair ranking frameworks in terms of producing better fairness evaluation scores. By separating fairness and relevance models, we utilized contextual features to construct tailored features for the fairness model. Compared with the existing fair ranking frameworks, which optimize fairness and relevance simultaneously, we better managed the fairness-relevance trade-off by adjusting the preference parameter with less computational cost. 

We showed that our DLF model improves fairness on initial rankings retrieved from BM25 \cite{Robertson1993} and RM3 \cite{lv2009comparative} and, ideally, more retrieval models. In the future, we will test more complex retrieval models, such as neural and dense retrieval models, to construct the initial rankings. Besides, we will further explore contextual features that can be used to improve leaning-based models, especially those features that bring more information regarding fairness and model interpretability, such as stance and sentiment. The leverage of contextual features also provides a new direction of data augmentation for future learning-based fair-ranking studies. Last, we are also interested in adjusting the weights shown in Eq.(\ref{final_loss}) to investigate whether weighting fairness categories differently helps overall fairness. 

\begin{acks}
This work is supported by the graduate fellowship from the Institute for Financial Services Analytics at University of Delaware. The authors thank all the ICTIR reviewers for their advice and suggestions.
\end{acks}

\bibliographystyle{ACM-Reference-Format}
\bibliography{sample-base}

\end{document}